\documentclass[preprint]{emulateapj}
\journalinfo{Submitted for publication in ApJ}
\submitted{Revised and and resubmitted for publication in ApJ, \today}
\usepackage{graphicx}
\usepackage{fancyref}
\usepackage{amsmath}
\usepackage{subfigure}
\usepackage{tabularx}
\usepackage{threeparttable}

\shorttitle{A New Model for Axion Halos from N-Body Simulations}
\shortauthors{Lentz, Quinn, Rosenberg, Tremmel}

\newcommand{\ngal}{{16} }
\newcommand{\alp}{${0.36 \pm 0.13}$}
\newcommand{\bet}{${1.39 \pm 0.28}$}
\newcommand{\T}{${(4.7 \pm 1.9) \times10^{-7}}$ }
\newcommand{\widthratio}{{1.8} }
\newcommand{\vctyp}{{226}}

\begin{document}

\title{ A New Signal Model for Axion Cavity Searches from N-Body Simulations}

\author{Erik W Lentz\altaffilmark{1}}
\author{Thomas R Quinn \altaffilmark{2}}
\author{Leslie J Rosenberg \altaffilmark{1}}
\author{Michael J Tremmel \altaffilmark{2}}

\altaffiltext{1}{Physics Department, University of Washington,
                 Seattle, WA 98195-1580;       
		 {\tt lentze@phys.washington.edu, ljrosenberg@phys.washington.edu}}
\altaffiltext{2}{Astronomy Department, University of Washington, 
                 Seattle, WA 98195-1580;       
		  {\tt trq@astro.washington.edu, mjt29@astro.washington.edu}}

\begin{abstract}

Signal estimates for direct axion dark matter searches have {used} the isothermal sphere halo model for the last several decades. While insightful, the isothermal model does not capture effects from a halo's infall history nor the influence of baryonic matter, which has been shown to significantly influence a halo's inner structure. The high resolution of cavity axion detectors can make use of modern cosmological structure-formation simulations, which begin from realistic initial conditions, incorporate a wide range of baryonic physics, and are capable of resolving detailed structure. This letter uses a state-of-the-art cosmological N-body+Smoothed-Particle Hydrodynamics simulation to develop an improved signal model for axion cavity searches. {Signal shapes} from a class of galaxies encompassing the Milky Way are found to depart significantly from the isothermal sphere. A new signal model for axion detectors is proposed and projected sensitivity bounds on {the} Axion Dark Matter eXperiment data are presented.

\end{abstract}


\section{Introduction}
\label{Introduction}
 
The axion is a compelling candidate for the dark matter (DM). The theory of axions originated in 1977 as a result of an axial symmetry over the QCD sector, used to solve the strong CP problem \citep{Peccei1977}. The following year, the axion particle was proposed from a posited spontaneous breaking of the axial symmetry at high energy scales \citep{Weinberg1978, Wilczek1978}. This initial PQWW axion was quickly ruled out (for a review of {axion experiment findings see \citet{PQWWruleout,PDG2016}}), giving way to models with growing application to cosmology \citep{Marsh2016}. The focus of this letter is to produce an updated signal model {useful} for QCD axions in terrestrial microwave cavity detectors based on cosmologically realistic simulations.

\begin{figure}[]
\begin{center}
\includegraphics[width=9cm]{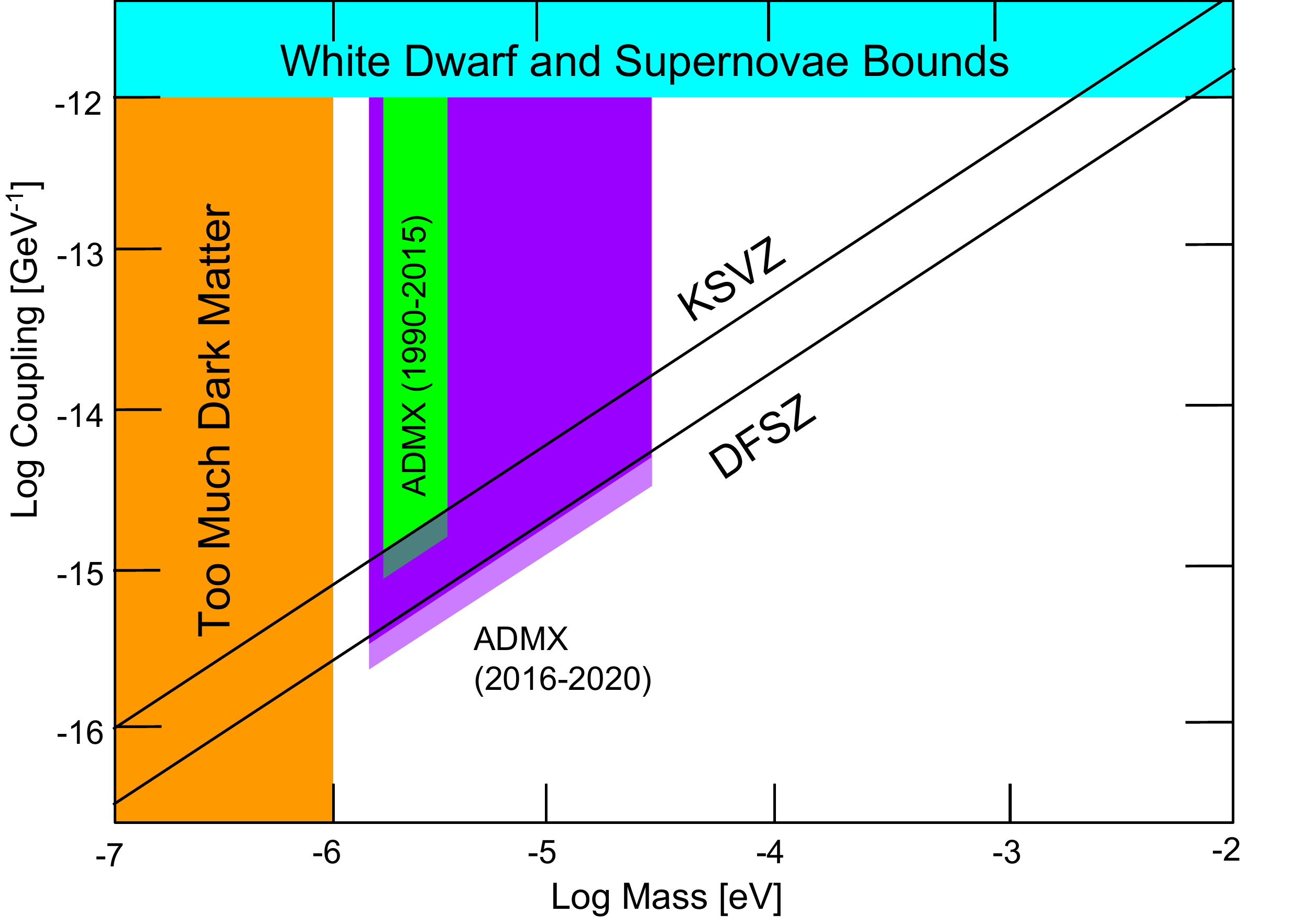}
\caption{Illustration of the QCD axion parameter space {over the plausible DM region,} with current astronomical and cosmological constraints, plus benchmark theories. The {90 \% confidence level} coupling bound from the Axion Dark Matter eXperiment's (ADMX's) Phase 1 operations through 2014 \citep{ADMX2014} is included in green, with { Generation 2 projections shown in purple \citep{PDG2016}}. Under the new signal model presented in Section \ref{Discussion}, an increase of the SNR of $\widthratio$ would translate to an improvement in the coupling limit of $\sqrt{\widthratio}$, illustrated in lighter shades.}
\label{fig:fig1}
\end{center}
\end{figure}

As a DM candidate, the QCD axion is highly attractive due to its well-bounded parameter space {of mass and coupling}, Fig. \ref{fig:fig1}. Starting at milli-eV masses, there is a bound above which the axion would have been seen in various astrophysical processes {\citep{Raffelt2008,Isern2010,Corsico2012,Viaux2013,PDG2016,Marsh2016}}. {Approaching} micro-eV masses, there is a bound below which the (misalignment) creation mechanism would produce more axions than there is dark matter {\citep{Abbott1983,Dine1983,Preskill1983,PDG2016}}. This lower bound is somewhat soft as axion creation mechanisms can be suppressed in the details of some axion theories \citep{Marsh2016}. The two diagonal lines represent benchmark axion models. KSVZ represents a theory where the axion couples to hadrons only \citep{Kim1979,Shifman1980}, and DFSZ couples to both hadrons and leptons \citep{Dine1981,Zhitnitsky1980} {as consistent with grand unified theories}. The search window is then given by the region between KSVZ and DFSZ and the lower and upper mass bounds. 

The axion's low mass {and feeble couplings} lead to unconventional search techniques. One attractive method used by axion DM searches threads a magnetic field through a resonant cavity, stimulating the decay of DM axions into microwaves via the effective CP-violating interaction
\begin{equation}
\mathcal{L}_{int} = \frac{g_{a \gamma \gamma}}{4 \pi} a F \tilde{F} 
\end{equation}
where $a$ is the axion field, $F$ and $\tilde{F}$ are the electromagnetic field strength and its dual, and $g_{a \gamma \gamma}$ is the coupling strength {\citep{Sikivie1983}}. The energy distribution of axions is sampled by tuning the cavity over large ranges in frequency, making the microwave power {spectrum} the figure of interest. The resonant microwaves are detected by a receiver sensitive to {sub-}yocto-watt power. {These experiments can have} frequency resolution to a part in $10^9$ or better \citep{ADMX2010}, well within the range of resolving fine structure in the axion distribution. {A relic axion density of $0.45$ GeV/cc is commonly used by cavity searches, motivated by Milky Way (MW) observations \citep{Gates1995}, making the focus of this letter to update the shape of a potential axion signal.}

Despite their \text{good} energy resolution, some axion searches use a filter {shape} derived from the Standard Halo Model (SHM) \citep{ADMX2010}, which comes from the assumption that the MW halo is given by a thermalized pressure-less self-gravitating sphere of particles. The velocity distribution of the SHM near the earth is given by a Maxwellian distribution below the escape velocity
\begin{equation}
f_{\vec{v}} \propto e^{- \vec{v}\cdot\vec{v}/ 2 \sigma_v^2}
\end{equation}
where {$\sigma_v = \sqrt{2} v_c $ \citep{BT2008} and the $v_c$ is the circular speed. To match the limits from the Axion Dark Matter eXperiment (ADMX) \citep{ADMX2010}, a value of $v_c =$ \vctyp km/s will be used, though recent observations put the solar orbital speed closer to $v_c = 255$ km/s \citet{Reid2014}.}The SHM is broadly predictive, but incapable of describing fine structure.

Since these searches {started}, there has been significant progress in simulating the formation of galaxies like the MW. One powerful simulation tool is the N-Body+Smoothed-Particle Hydrodynamics (N-Body+SPH) method. Capable of accurately resolving the inner structures of galaxies and their halos using cold dark matter and well-calibrated baryonic models, N-Body+SPH simulations are poised to give accurate models of DM structure for direct searches. This has already been done for WIMPs, which have not yielded significant differences from the SHM \citep{Sloane2016,Kelso2016,Bozorgnia2016}. As the energy spectra for cavity axion searches is different from the speed spectra relevant to WIMP nuclear recoil searches, an axion-specific analysis of structure-formation simulations is worthwhile.

This letter presents an axion-specific signal study using the recent Romulus25 (R25) N-Body+SPH simulation generated by the N-Body Shop \citep{Tremmel2016}, and proceeds as follows: Section \ref{Methods} presents R25 and the software tools utilized to select halos similar to the MW; Section \ref{Results} compares the selected halos with previous work and presents the axion-relevent spectra; Section \ref{Discussion} discusses how the results compare to the SHM and their implications to axion searches; Section \ref{Conclusions} consolidates findings and proposes how they could be used in ongoing axion searches.

\section{Methods}
\label{Methods}

This letter utilizes the results of the R25 N-Body+SPH simulation produced by the UW N-Body Shop \citep{Tremmel2016}, based on the ChaNGa N-Body+SPH code. R25 was created on the Blue Waters petascale computing facility. To analyze specific galaxies and halos in R25, the Amiga Halo Finder (AHF) \citep{Knollmann2009} is used to create the catalogue on R25, also at Blue Waters. Finally, the analysis was greatly assisted by the Pynbody N-body analysis software package \citep{Pontzen2013}.

\subsection{Romulus25}

R25 describes a $25$ Mpc periodic cosmological box filled with DM particles, evolving gas and star particles, and super-massive black holes (SMBHs). The box is placed in a $\Lambda$CDM cosmological setting according to findings from Planck ($\Omega_0=0.3086$, $\Lambda=0.6914$, $h=0.67$, $\sigma_8=0.77$; \citep{Planck2015}). Particle masses of $3.39 \times 10^5 M_{\odot}$ and $2.12 \times 10^5 M_{\odot}$ are used for the DM  and gas respectively. DM is over-sampled by a factor of $3.375$ relative to the gas, producing more accurate SMBH dynamics \citep{Tremmel2015,Tremmel2016}, and will reduce numerical noise in the analysis. The ChaNGa N-Body+SPH code uses a Barnes-Hut tree to calculate gravity with hexadecapole expansion of nodes. Time-stepping is done with a leapfrog integrator with individual time-steps for each particle. In addition to gravity, calibrated baryonic physics including star formation and evolution, cosmic UV background, supernovae feedback, primordial cooling, and SPH are used to govern the gas and stars. The SMBHs in R25 are implemented with realistic formation, dynamical friction-driven evolution, and accretion principles governed by local gas dynamics \citep{Tremmel2015,Tremmel2016}. A Plummer equivalent force-softening length of $250$ pc is used, of similar size to \citet{Sloane2016}, which is more than sufficient to resolve the local axion distribution on the kpc scale. The system is evolved to $z=0$ with a force accuracy/node opening criterion $\theta=0.65$ until $z=2$, after which an opening of $\theta=0.9$ is used. The time-stepping accuracy used is such that the {time step} $\Delta t < \eta \sqrt{\epsilon/a}$, where $\epsilon$ is the gravitational softening, $a$ is the acceleration of a particle, and $\eta$ is an accuracy criterion; $\eta =  0.185$ is used. ChaNGa is part of the AGORA \citep{Kim2014} code comparison collaboration. The code base has been thoroughly tested and has contributed to many astronomical topics including DM candidate testing \citep{Purcell2011,Menon2015,Fry2015a,Kim2016,Tomozeiu2016,Pontzen2016,Tremmel2016,Anderson2016}
\footnote{This letter uses a customized version of the ChaNGa code. A public distribution of ChaNGa is available via the UW N-Body Shop GitHub page (https://github.com/N-BodyShop/changa). Literature on its operation can be found on the wiki page (https://github.com/N-BodyShop/changa/wiki/ChaNGa).}.

R25 is large enough that it {has} O(30) MW-mass halos at z=0. Such a large set allows for a sampling of galaxies where reduction down to a MW-like sample is quantifiable via the use of filters on a halos catalogue.

\subsection{Halo Catalogues}

The AHF cataloging program is used to identify halos within the simulation. It operates under the principle of identifying halo centers via density-triggered domain refinement, collection of a halo's domain through successive inclusion of particle shells, and removal of particles unbounded to the halo \citep{Knollmann2009}. Beyond finding halos at a single timestep, AHF is capable of tracing individual halos through time and absorption in order to construct a merger tree. Through this tree, quantities like time of last major merger and more detailed merger history can be calculated, which will be important for the selection of halo-galaxy structures that resemble the MW. The TANGOS database package \citep{Pontzeninprep} is also utilized over the catalogue to aid in the selection process.

\subsection{Filters}

The galaxies analyzed here are selected using filters in line with the current understanding of MW structure and formation history \citep{MW1,MW2,MW3} without being so constraining as to limit halo statistics.
\begin{align}
& 0.5\times10^{12} M_{\odot} \le M_{vir} \le 1.6 \times10^{12} M_{\odot} \label{mass} \\
& R_{vir} \le 250 \text{ kpc} \\
& z_{major} \ge 0.75 \label{merger} \\
& 175 \text{ km/s} \le v_{circ} \le 275 \text{ km/s} \label{vcirc}
\end{align}
where $M_{vir}$ is the virialized mass, $R_{vir}$ is the enclosing virial radius, and $z_{major}$ is the redshift of last major merger, which is set by the progenitor ratio of 4:1. Several halos experienced no major mergers during the simulation.  $v_{circ}$ is the circular velocity in the plane of the galaxy at $8$ kpc, which did eliminate several halos from the set satisfying Eqs. \ref{mass} - \ref{merger}. At $z=0$, R25 contains \ngal halos which satisfy the filters; Table \ref{table:table1} contains some characteristics of each.

\begin{table}[]
\centering
\title{}
\caption{Relevant MW-like Galaxy Properties}
\begin{tabular}{llll}
\hline
\multicolumn{1}{|l|}{halo \#} & \multicolumn{1}{l|}{$M_{vir}$ ($10^{12} M_{\odot}$)} & \multicolumn{1}{l|}{$v_{circ}$ (km/s)} & \multicolumn{1}{l|}{$z_{major} $}   \\ \hline
\multicolumn{1}{|l|}{32} & \multicolumn{1}{l|}{$1.56$} & \multicolumn{1}{l|}{{202}} & \multicolumn{1}{l|}{1.06}   \\ \hline
\multicolumn{1}{|l|}{33} & \multicolumn{1}{l|}{$1.42$} & \multicolumn{1}{l|}{{212}} & \multicolumn{1}{l|}{3.00}   \\ \hline
\multicolumn{1}{|l|}{34} & \multicolumn{1}{l|}{$1.39$} & \multicolumn{1}{l|}{{226}} & \multicolumn{1}{l|}{2.38}  \\ \hline
\multicolumn{1}{|l|}{36} & \multicolumn{1}{l|}{$1.15$} & \multicolumn{1}{l|}{{209}} & \multicolumn{1}{l|}{8.00}  \\ \hline
\multicolumn{1}{|l|}{37} & \multicolumn{1}{l|}{$1.03$} & \multicolumn{1}{l|}{{248}} & \multicolumn{1}{l|}{4.59}  \\ \hline
\multicolumn{1}{|l|}{{39}} & \multicolumn{1}{l|}{${1.00}$} & \multicolumn{1}{l|}{{226}} & \multicolumn{1}{l|}{ {1.69}}  \\ \hline
\multicolumn{1}{|l|}{42} & \multicolumn{1}{l|}{$0.98$} & \multicolumn{1}{l|}{{178}} & \multicolumn{1}{l|}{5.45}  \\ \hline
\multicolumn{1}{|l|}{44} & \multicolumn{1}{l|}{$0.92$} & \multicolumn{1}{l|}{{194}} & \multicolumn{1}{l|}{--}   \\ \hline
\multicolumn{1}{|l|}{46} & \multicolumn{1}{l|}{$0.77$} & \multicolumn{1}{l|}{{207}} & \multicolumn{1}{l|}{3.41}   \\ \hline
\multicolumn{1}{|l|}{47} & \multicolumn{1}{l|}{$0.73$} & \multicolumn{1}{l|}{{222}} & \multicolumn{1}{l|}{5.99}   \\ \hline
\multicolumn{1}{|l|}{48} & \multicolumn{1}{l|}{$0.71$} & \multicolumn{1}{l|}{{195}} & \multicolumn{1}{l|}{2.25}   \\ \hline
\multicolumn{1}{|l|}{{50}} & \multicolumn{1}{l|}{{${0.66}$}} & \multicolumn{1}{l|}{{199}} & \multicolumn{1}{l|}{{3.09}}   \\ \hline
\multicolumn{1}{|l|}{51} & \multicolumn{1}{l|}{$0.77$} & \multicolumn{1}{l|}{{185}} & \multicolumn{1}{l|}{--}   \\ \hline
\multicolumn{1}{|l|}{53} & \multicolumn{1}{l|}{$0.55$} & \multicolumn{1}{l|}{{207}} & \multicolumn{1}{l|}{2.52}   \\ \hline
\multicolumn{1}{|l|}{55} & \multicolumn{1}{l|}{$0.63$} & \multicolumn{1}{l|}{{195}} & \multicolumn{1}{l|}{2.01}   \\ \hline
\multicolumn{1}{|l|}{60} & \multicolumn{1}{l|}{$0.55$} & \multicolumn{1}{l|}{{176}} & \multicolumn{1}{l|}{--}   \\ \hline
\end{tabular}
\begin{tablenotes}
      \small
      \item Selected parameters of the simulated MW-like galaxies analyzed in this paper: $M_{vir}$ is the total virial mass of the halo and galaxy, $v_{circ}$ is the {in-disk} circular velocity at $8$ kpc from the center of the galaxy, $z_{major}$ is the redshift of the last major merger, where one exists.
\end{tablenotes}\label{table:table1}
\end{table}

\section{Results}
\label{Results}

\begin{figure*}[h]
\centering
    \includegraphics[width=0.47\linewidth]{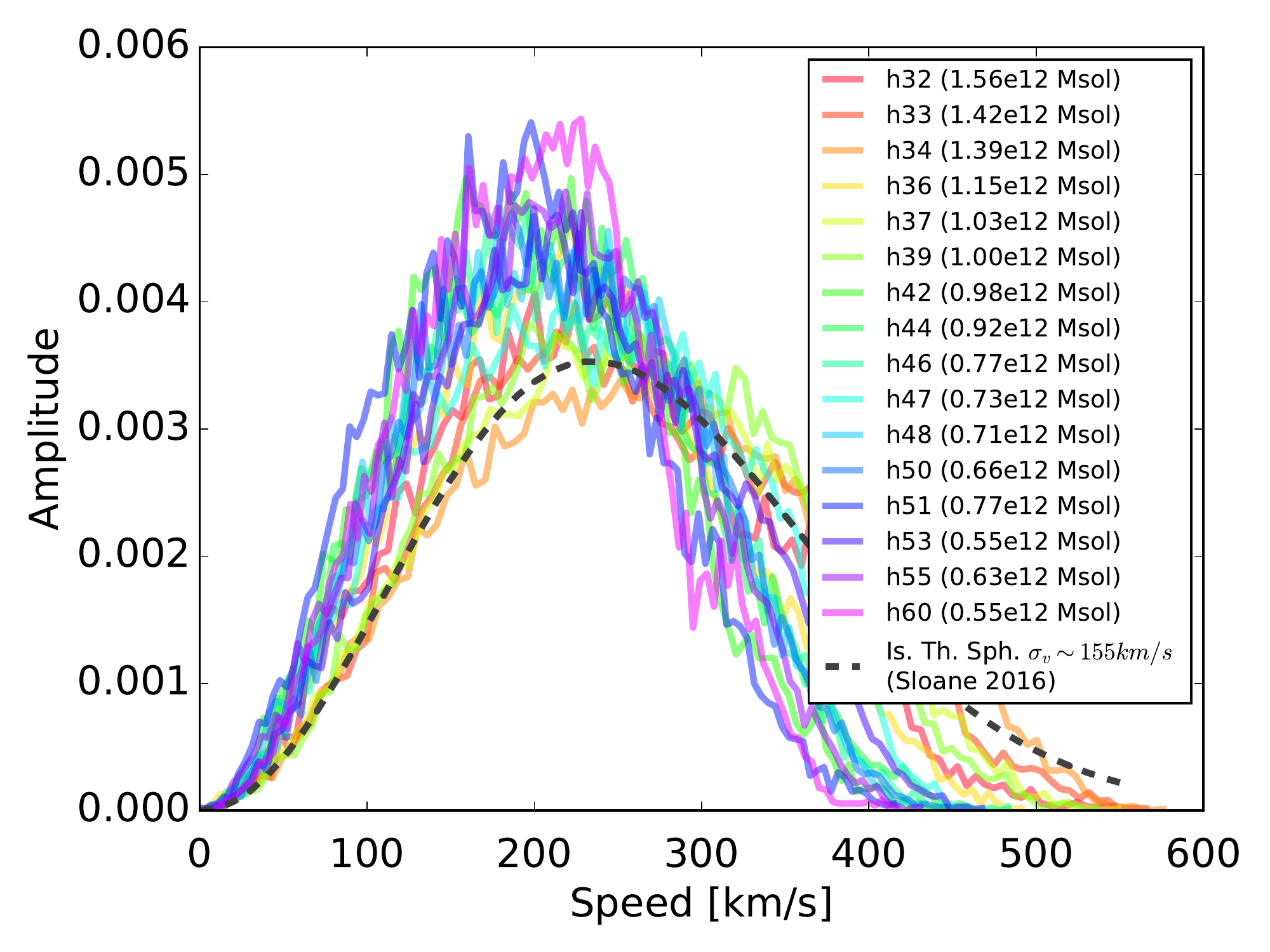}\hfil
    \includegraphics[width=0.47\linewidth]{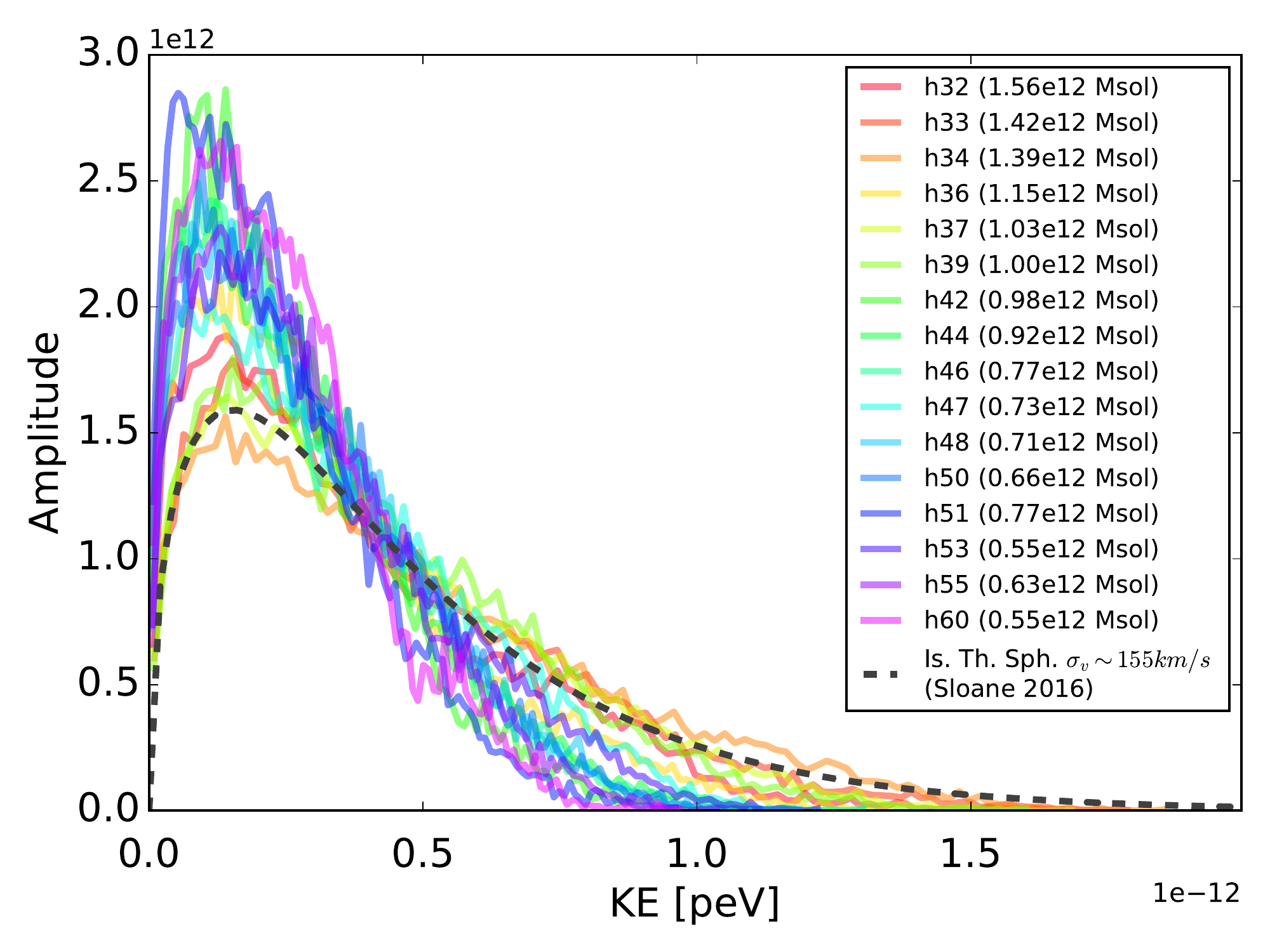}\par\medskip
    \includegraphics[width=0.47\linewidth]{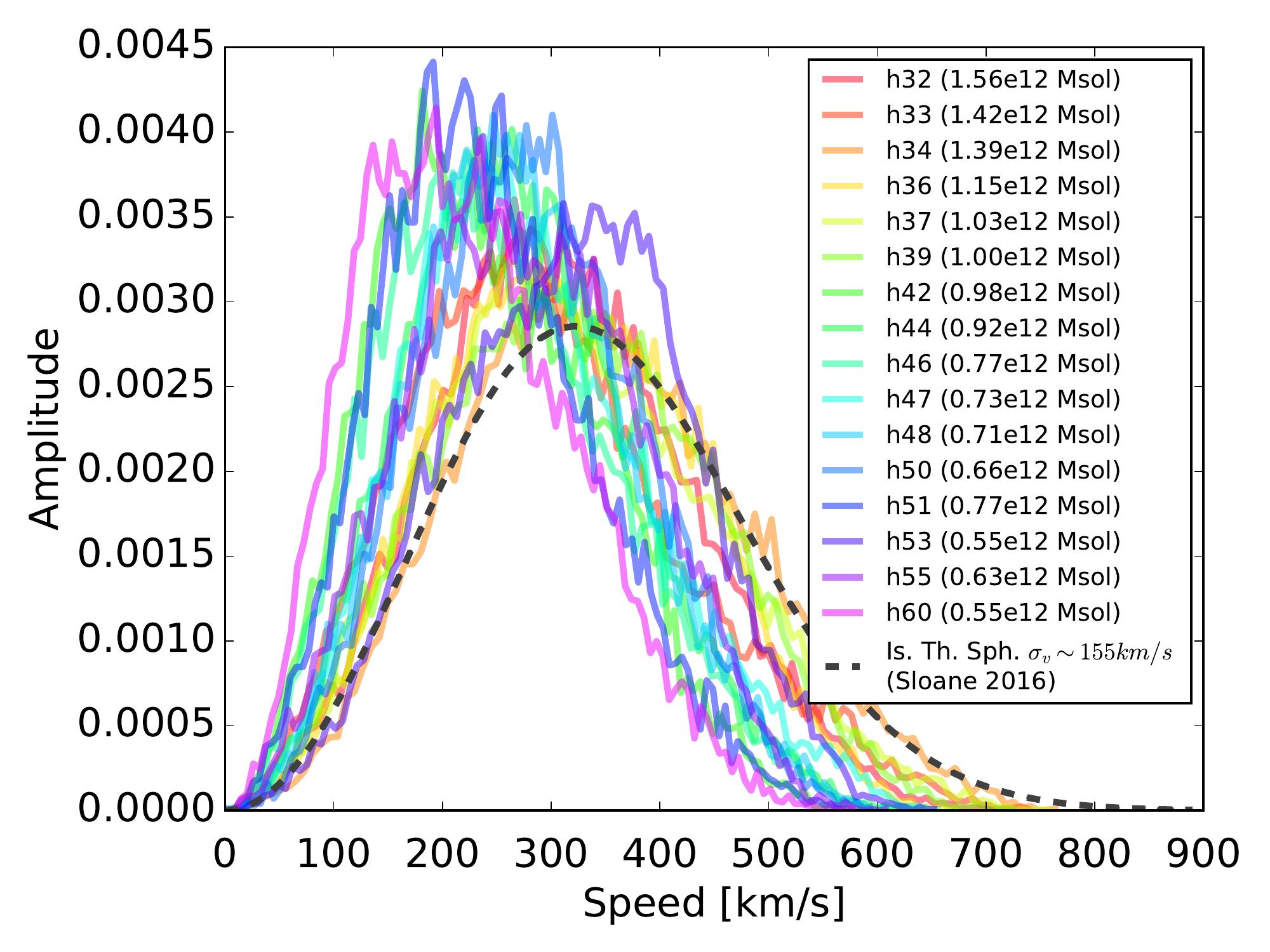}\hfil
    \includegraphics[width=0.47\linewidth]{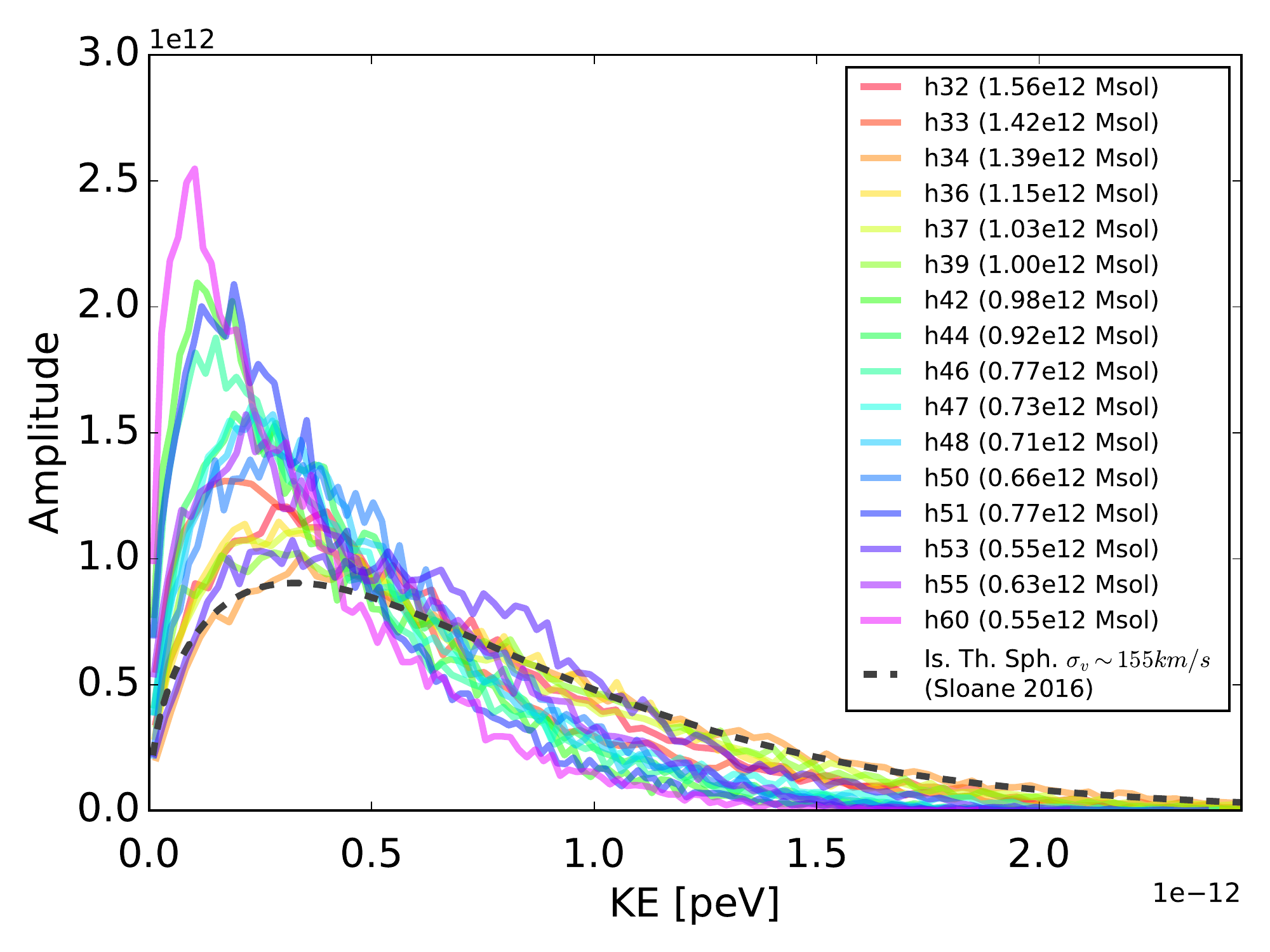}
\caption{Spectra of MW-like halos from Romulus25 at $z=0$, normalized to unity. The spectra are taken from galaxy-centered in-disk toroidal samples of cylindrical extent $-2 kpc \le z \le 2 kpc$, $r_l-1 kpc \le r \le r_l+1 kpc$, where $r_l = 8$ kpc, though the results are fairly robust to changes in the value of $r_l$. The galaxy center frame speed (upper left) and energy spectra (upper right) show general agreement with the SHM, with small departures in the low-speed spectra amplified in the energy spectra, primarily due to the shift from a measure linear in particle speed to quadratic. The solar frame speed spectra (lower left) shows additional low-speed departures from the SHM, but are still consistent with previous WIMP studies (e.g., \citet{Sloane2016, Schaller2016}). The solar frame energy spectra (lower right) displays drastic departure from the SHM, which is expected due to the change in measure when transforming from speed to energy coordinates. The SHM shape in the galactic frame have been boosted {tangentially by $\bar{v}_c = \vctyp$ km/s, in line with ADMX analyses \citep{ADMX2010}.}}
\label{fig:fig3}
\end{figure*}

Before presenting the findings of the axion signal analysis, it is prudent to first establish the sensibility of MW-like halos in R25. Fig. \ref{fig:fig3} shows that the speed spectra of solar-radius samples in the galactic frame are consistent with Figures 1 \& 2 in {\citet{Sloane2016}} and Figure 1 in {\citet{Schaller2016}}, and are well fit by the SHM. {The halos' densities at the solar radius are also seen to match results from \citet{Sloane2016} and \citet{Bozorgnia2016} over their respective mass ranges.} The solar sample of particles is given by a 2 kpc by 4 kpc toroid about the solar {orbit, which is taken to be in the baryonic disk,}
\begin{equation}
-2 \text{ kpc} \le z \le 2 \text{ kpc} \text{ ,  } r_l-1 \text{ kpc} \le r \le r_l+1 \text{ kpc} \label{torus_dim}
\end{equation}
where $r_l$ is chosen to match the MW solar radius of $8$ kpc, though the spectral shapes are reasonably robust to the choice of radius. Sample sizes for these toroids ranged from 11,000 to 28,000 DM particles. 

The {galactic frame} energy spectra of Fig. \ref{fig:fig3} also shows a {similar} consistency with the SHM, though small speed departures become amplified due to changes in measure from speed to energy.  

Terrestrial DM search experiments are moving with respect to the galactic center. The lab is held to be in the galaxy's local standard of rest, approximated as a circular orbit about the center of the galaxy of $r_l = 8$ kpc, coincident with the Sun-MW orbital shape and radius. This seemingly trivial choice of radius for galaxies other than the MW does not affect the structure of the resulting spectra, which are found to be robust over a span of $6 \text{ kpc} \le r_l \le 18 \text{ kpc}$. The orbit speed of the lab is given by setting the acceleration of disk particles at the orbit radius to orbital motion
\begin{equation}
\bar{a}(r_l) = \frac{v_l^2}{r_l}
\end{equation}
where the RHS is the centripetal acceleration of the lab frame and $\bar{a}$ is the average acceleration at the lab radius. In order to approximate the velocity distribution in a point-like lab, this sample region assumes a cylindrically symmetric, homogeneous, equi-potential, steady-state system within. To evaluate the particles in the lab frame, a Galilean boost is performed on each sample particle using the solar orbital velocity
\begin{equation}
(v_r,v_t,v_z) \to (v_r,v_t - v_l,v_z) \label{lab_boost}
\end{equation}
where $v_l$ is taken to be the circular velocity in the presence of DM+baryons at the orbit radius and $v_r$, $v_t$, $v_z$ are a particle's radial, tangential, and z-component velocities respectively. The laboratory speed, energy, and other spectra can now be calculated by forming distribution functions over the desired observable. 

The laboratory-frame speed spectra of the R25 halos shown in Fig. \ref{fig:fig3} display some deviations from the SHM, particularly in the low-speed region, which suggests bulk motion. Prior WIMP DM studies also observe these deviations, though they find more pronounced low-speed excesses in DM-only simulations as opposed to when baryons are present. The galactic and solar frame speed distributions, including baryons, are consistent with the count-rate proxy of integrated weighted velocity functions, $g(v_{min})$, of \citet{Sloane2016}. 

In contrast, the {lab frame} energy spectra show a marked difference from the SHM, Fig. \ref{fig:fig3}. The deviation towards the lower relative speed becomes much more pronounced due to the change in measure from $dm \sim dv  $ to $dm {\sim dE} \sim v dv $. This apparent narrowing and enhancement of the signal is {present over all but one of the} \ngal halos. {The degree of narrowing is robust over the halos, with analyses of signal width over local DM density and halo mass showing no significant correlations.} 

{Also, recall that filters in Eqs. \ref{mass} - \ref{vcirc} allow galaxies which deviate from the MW. Specifically, the galaxies inside halos h32, h34, h36, h44, and h48 have older stars and larger bulges than is expected for the MW. These galaxies lower the mean deviation as they produce spectra preferentially closer to the SHM than the rest of the halos; such halos are still included, as filters based on morphology are not considered here.}

\section{Discussion}
\label{Discussion}

\begin{figure}[]
\begin{center}
\includegraphics[width=8.5cm]{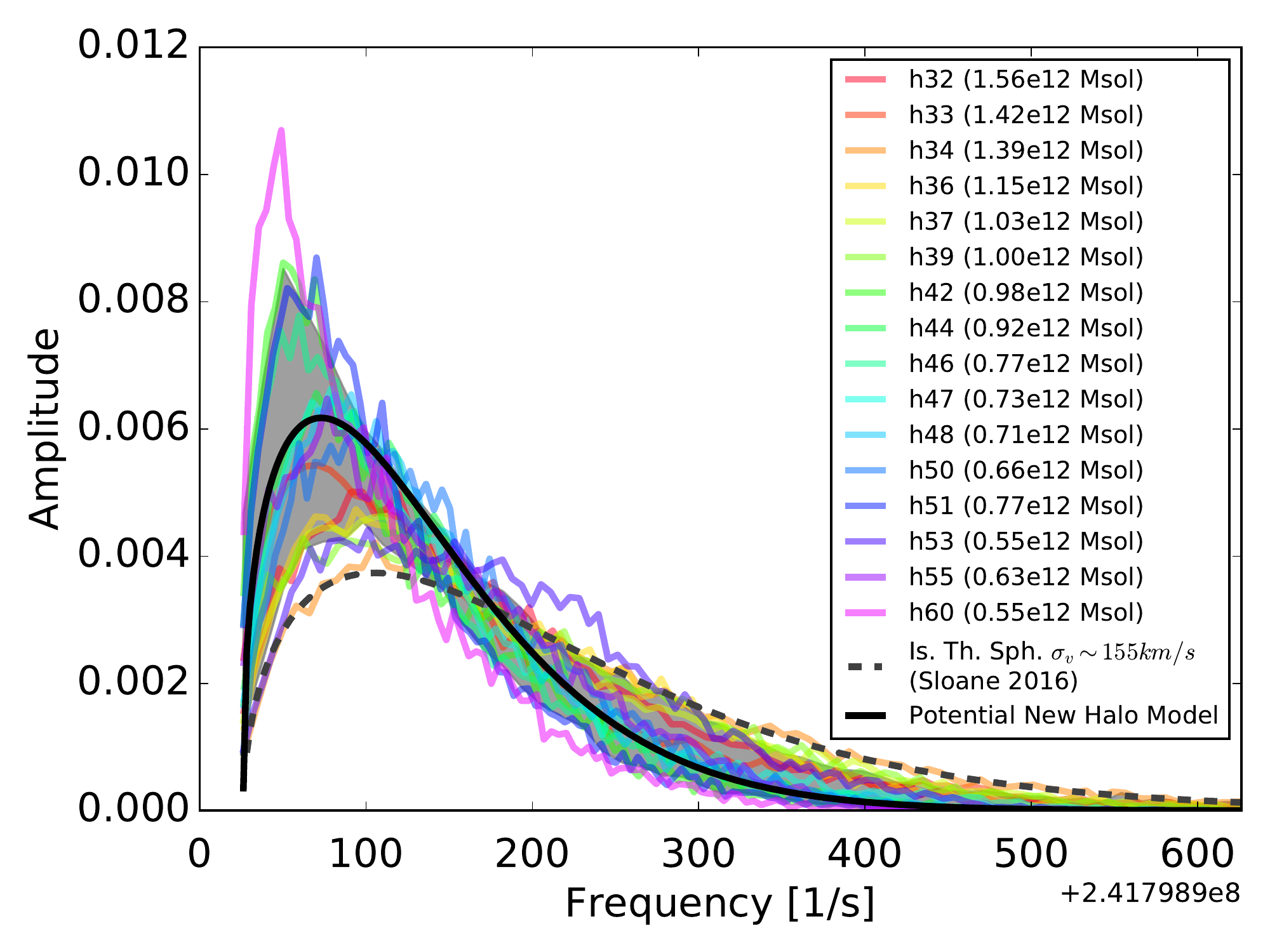}
\caption{Frequency spectra of MW-like halos from Romulus25 at $z=0$ and the SHM composed of $10^{-6} eV$ axions, generated from Fig. \ref{fig:fig3} spectra via the energy-to-frequency transform derivable from Eq. \ref{en_eq}. The solid black line represents the new {shape} of the form Eq. \ref{new_model} fitted to to the halos, with the gray representing the data-based error estimate {using the two-thirds rule.}}
\label{fig:fig7}
\end{center}
\end{figure}

There is a significant difference between galactic and lab frames in the spectra. The lab-frame energy spectra are sensitive to bulk motions, which contribute to the signal narrowing, though there is not yet a full understanding of the mechanisms behind this effect. The characteristic spectral shape over a wide range of galaxies communicates a level of robustness in the line shape and gives confidence in a signal model based on these spectra. The signal shapes expected to be observed by a microwave cavity experiment in each R25 halo are calculated, shown in Fig. \ref{fig:fig7} for a $10^{-6} eV$ axion. The conversion to frequency uses the equivalence between the energy of the decayed axion and the resonant microwave's energy
\begin{equation}
E_{\gamma} = \hbar \nu = m_a c^2 + m_a v^2/2 \label{en_eq}
\end{equation}
where $m_a$ is the axion mass, $v$ is the axion speed in the lab frame, and $\nu$ is the microwave frequency. The solid black line of Fig. \ref{fig:fig7} represents the fitted model signal shape. The proposed signal shape keeps a Maxwellian-like form
\begin{equation}
f_{\nu} \propto \left( \frac{(\nu - \nu_o) h}{mT} \right)^{\alpha} e^{-\left( \frac{(\nu - \nu_o) h}{mT} \right)^{\beta}} \label{new_model}
\end{equation}
{where the parameters are constrained to be positive. The best-fit parameters are found using a log-normal local M-estimate and calculated to be $\alpha =$ \alp, $\beta =$ \bet, and $T =$ \T, with the errors given by projections of the covariance matrix.}

The narrowing of the observed shapes has implications for axion search experiments. The modeled signal has a {90}\% width---the minimum span which contains {90}\% of the distribution---that is \widthratio times narrower than the SHM. Such a narrowing {of the signal shape} would improve a search's SNR by the same factor. For an axion cavity search like ADMX, the increase in sensitivity translates to an improvement in the coupling limit of $\sqrt{\widthratio}$,{ suggesting past data runs \citep{ADMX2010,ADMX2014} have near-DFSZ} sensitivity, Fig. \ref{fig:fig1}.

\section{Conclusions}
\label{Conclusions}

This letter demonstrates that halos from the Romulus25 simulation show major differences from the SHM in observables relevant to axion DM searches. The class of  galaxies satisfying Eqs. \ref{mass} - \ref{vcirc}, which {include} the MW, are observed to produce significantly narrower DM energy spectra than the SHM. This narrowing may be caused by bulk rotational motions or other biases toward slow motion relative to the baryonic disk. While beyond the scope of this paper, an explanation for the distribution shape may lie with {merger history, }the influence of the baryonic disk and the incompleteness of gravitational virialization. 

The Romulus25 simulation provides a means to {predict realistic} axion signals. This high-resolution simulation develops realistic disk galaxies using robust gas dynamics and star formation/feedback models, which reproduce observed galaxy properties \citep{Tremmel2016,Governato2007}. Further, Romulus25's large size and uniform resolution provide a statistically significant sample of galaxies similar to the MW, and produce a consistent spectral shape in the circularly-orbiting frame. Together, the updated signal model {from Eqn. }\ref{new_model} provides a marked improvement in the signal shape for axion cavity searches. {A conservative estimate of the new shape} would serve to improve the signal-to-noise of axion cavity searches by $\widthratio$, increasing the sensitivity of previous {analyses} and improving future {observations} through the option of {scanning} at a higher sensitivity or a higher search rate. Such an optimization would result in faster and more sensitive searches.

\section{Acknowledgements}
\label{Acknowledgements}

{We would like to thank the referee for the productive discussion in the refinement of this paper. We also} gratefully acknowledge the support of the U.S. Department of Energy office of High Energy Physics and the National Science Foundation. TQ and MT were supported in part by the NSF grant AST-1514868. EL and LR were supported by the DOE grant DE-SC0011665. This research is part of the Blue Waters sustained-petascale computing project, which is supported by the National Science Foundation (awards OCI-0725070 and ACI-1238993) and the state of Illinois. Blue Waters is a joint effort of the Univeristy of Illinois at Urbana-Champaign and its National Center for Supercomputing Applications. This work is also part of a PRAC allocation support by the National Science Foundation (award number OCI-1144357).

\end{document}